\documentstyle[12pt,epsfig]{article}

\newcommand{\agt}{\,\rlap{\lower 3.5 pt \hbox{$\mathchar \sim$}} \raise 1pt
 \hbox {$>$}\,}
\newcommand{\alt}{\,\rlap{\lower 3.5 pt \hbox{$\mathchar \sim$}} \raise 1pt
 \hbox {$<$}\,}

\begin{document}

\title{\vskip-3cm{\baselineskip14pt
\centerline{\normalsize DESY 99-124\hfill ISSN 0418-9833}
\centerline{\normalsize KEK-TH-640\hfill}
\centerline{\normalsize hep-ph/9909517\hfill}
\centerline{\normalsize September 1999\hfill}
}
\vskip1.5cm
Massive-Evolution Effects on Charmonium Hadroproduction
}
\author{Bernd A. Kniehl$^{1,}$\thanks{Permanent
address: II. Institut f\"ur Theoretische Physik, Universit\"at Hamburg,
Luruper Chaussee 149, 22761 Hamburg, Germany.}
\ and Lennart Zwirner$^2$\\
{\normalsize $^1$
High Energy Accelerator Research Organization (KEK), Theory Division,}\\
{\normalsize 1-1 Oho, Tsukuba-shi, Ibaraki-ken, 305-0801 Japan}\\
{\normalsize $^2$
II. Institut f\"ur Theoretische Physik, Universit\"at Hamburg,}\\
{\normalsize Luruper Chaussee 149, 22761 Hamburg, Germany}}

\date{}

\maketitle

\thispagestyle{empty}

\begin{abstract}
The fragmentation functions $D_{a\to H}(x,\mu^2)$ of a heavy hadron $H$, with
mass $m_H$, satisfy the phase-space constraint $D_{a\to H}(x,\mu^2)=0$ for
$x<m_H^2/\mu^2$, which is violated by the naive $\mu^2$ evolution equations.
Using appropriately generalized $\mu^2$ evolution equations, we reconsider the
inclusive hadroproduction of prompt $J/\psi$ mesons with high transverse momenta
in the framework of the factorization formalism of nonrelativistic quantum
chromodynamics, and determine the resulting shifts in the values of the leading
colour-octet matrix elements, which are fitted to data from the Fermilab
Tevatron. 

\medskip

\noindent
PACS numbers: 13.85.-t, 13.85.Ni, 13.87.Fh, 14.40.Gx
\end{abstract}

\newpage

In the framework of the QCD-improved parton model, the inclusive production of
single hadrons is described by using fragmentation functions,
$D_{a\to h}(x,\mu^2)$.
The value of $D_{a\to h}(x,\mu^2)$ corresponds to the probability for a parton
$a$ which comes out of the hard-scattering process to form a jet which contains
a hadron $h$ carrying the longitudinal-momentum fraction
$x=(p_h^0+p_h^3)/(p_a^0+p_a^3)$, where $p_a$ and $p_h$ are the four-momenta of
$a$ and $h$ in the infinite-momentum frame.
Here, $\mu$ is the fragmentation scale, which is typically chosen to be of the
order of the centre-of-mass energy $\sqrt s$ (transverse momentum $p_T$ of $h$)
in lepton-lepton (lepton-hadron and hadron-hadron) collisions.
At next-to-leading order (NLO), $\mu$ is identified with the mass scale at which
the collinear singularities associated with the outgoing parton $a$ are
factorized.
In the case of a light hadron, with mass $m_h\ll\mu$, the $\mu^2$ dependence of
$D_{a\to h}(x,\mu^2)$ is determined by the timelike Altarelli-Parisi evolution
equations,
\begin{equation}
\frac{\mu^2\partial}{\partial\mu^2}D_{a\to h}(x,\mu^2)
=\sum_b\int_x^1\frac{dy}{y}P_{a\to b}^{(T)}(y,\mu^2)
D_{b\to h}\left(\frac{x}{y},\mu^2\right),
\label{eq:ap}
\end{equation}
where $P_{a\to b}^{(T)}(x,\mu^2)$ are the timelike $a\to b$ splitting functions.
Ready-to-use expressions for $P_{a\to b}^{(T)}(x,\mu^2)$ through NLO are
collected in the Appendix of Ref.~\cite{bin}.\footnote{%
There is an obvious typographical error in the published version of
Ref.~\cite{bin}, which was absent in the preprint version thereof.
In the line before the last of Eq.~(17), $\ln\ln(1-x)$ should be replaced by
$\ln(1-x)$.}

In Eq.~(\ref{eq:ap}), $x$ may, in principle, be arbitrarily low for any value of
$\mu$.
In the case of a heavy hadron $H$, with mass $m_H\alt\mu$, this conflicts with
the fact that the fragmentation process is only allowed by kinematics if the
virtuality $p_a^2$ of the fragmenting parton $a$ satisfies the condition
$p_a^2>m_H^2/x$.
This may be understood as follows.
Consider the general situation where $a$ fragments into $n$ particles, with
masses $m_i$ and four-momenta $p_i$, and choose the coordinate system so that
$p_a^\mu=\left(p_a^0,0,0,p_a^3\right)$.
Defining the longitudinal-momentum fractions as
$x_i=(p_i^0+p_i^3)/(p_a^0+p_a^3)$, we then have
\begin{equation}
p_a^2=\left(p_a^0+p_a^3\right)\sum_{i=1}^n\left(p_i^0-p_i^3\right)
=\sum_{i=1}^n\frac{m_i^2+p_{T,i}^2}{x_i},
\end{equation}
where $p_{T,i}=\sqrt{\left(p_i^1\right)^2+\left(p_i^2\right)^2}$ are the
intrinsic transverse momenta.
If all final-state particles, except for $H$, are light and the intrinsic
transverse momenta are neglected, as is usually done in the parton model, then
it follows that $p_a^2\ge m_H^2/x$.
If we identify $\mu^2=p_a^2$, then this phase-space constraint leads to the
condition $D_{a\to H}(x,\mu^2)=0$ for $x<m_H^2/\mu^2$.

In order to properly implement this condition, Eq.~(\ref{eq:ap}) must be
generalized.
Such a generalization was proposed in Refs.~\cite{bra,man} for the $\mu^2$
evolution of $D_{g\to H}(x,\mu^2)$ generated by the $g\to g$ splitting.
It is straightforward to extend this formalism to also include the fragmentation
of the other partons and the nondiagonal evolution effects.
This leads us to the ansatz
\begin{equation}
D_{a\to H}(x,\mu^2)=\sum_b\int_{m_H^2}^{\mu^2}\frac{dq^2}{q^2}\int_x^1\frac{dy}{y}
G_{a\to b}(y,q^2;\mu^2)d_{b\to H}\left(\frac{x}{y},q^2\right),
\label{eq:d}
\end{equation}
where $G_{a\to b}(y,q^2;\mu^2,)$ corresponds to the probability for the parton
$a$, with virtuality $\mu^2$, to emit a parton $b$, with longitudinal-momentum
fraction $y$ and virtuality $q^2$, and $d_{b\to H}(z,q^2)$ to the one for parton
$b$ to subsequently decay to a hadron $H$, with longitudinal-momentum fraction
$z$ relative to $b$.
The $\mu^2$ dependence of $G_{a\to b}(y,q^2;\mu^2)$ is determined by the
evolution equation
\begin{equation}
\frac{\mu^2\partial}{\partial\mu^2}G_{a\to b}(y,q^2;\mu^2)
=\sum_c\int_y^1\frac{dz}{z}P_{a\to c}^{(T)}(z,\mu^2)
G_{c\to b}\left(\frac{y}{z},q^2;z\mu^2\right),
\label{eq:g}
\end{equation}
which is similar to Eq.~(\ref{eq:ap}), except that the virtuality of the
intermediate parton $c$ is taken to be $z\mu^2$ instead of $\mu^2$.\footnote{%
Here, we deviate from Eq.~(4) of Ref.~\cite{bra}, which refers to the evolution
in $q^2$.}
Furthermore, $G_{a\to b}(y,q^2;\mu^2)$ satisfies the boundary condition
\begin{equation}
G_{a\to b}(y,\mu^2;\mu^2)=\delta_{ab}\delta(1-y).
\label{eq:b}
\end{equation}
According to the above argument, $G_{a\to b}(y,q^2;\mu^2)$ is subject to the
phase-space constraint $G_{a\to b}(y,q^2;\mu^2)=0$ for $y<q^2/\mu^2$.
For the same reason, we have $d_{b\to H}(z,q^2)=0$ for $z<m_H^2/q^2$.
From Eq.~(\ref{eq:d}) it hence follows that $D_{a\to H}(x,\mu^2)=0$ for
$x=yz<m_H^2/\mu^2$ as desired.

In the perturbative calculation of the fragmentation function
$D_{a\to H}\left(x,\mu_0^2\right)$ at the initial scale $\mu_0$, with
$\mu_0\agt m_H$, $d_{a\to H}(x,q^2)$ acts as a source density, in the sense that
\begin{equation}
D_{a\to H}\left(x,\mu^2_0\right)=\int_{m_H^2}^{\mu_0^2}\frac{dq^2}{q^2}
d_{a\to H}(x,q^2),
\label{eq:i}
\end{equation}
which follows from Eq.~(\ref{eq:d}) by approximating $G_{a\to b}(y,q^2;\mu^2)$
with the aid of Eq.~(\ref{eq:b}).
In practice, the upper bound of integration in Eq.~(\ref{eq:i}) is taken to be
infinity because the integrand rapidly falls off with increasing $q^2$
\cite{psi,chi}.

Differentiating Eq.~(\ref{eq:d}) with respect to $\ln\mu^2$ and substituting
Eqs.~(\ref{eq:g}) and (\ref{eq:b}) on the right-hand side, we may eliminate
$G_{a\to b}(y,q^2;\mu^2)$ and thus obtain a single set of inhomogenious
integro-differential evolution equations for $D_{a\to H}(x,\mu^2)$, namely,
\begin{equation}
\frac{\mu^2\partial}{\partial\mu^2}D_{a\to H}(x,\mu^2)=d_{a\to H}(x,\mu^2)+
\sum_b\int_x^1\frac{dy}{y}P_{a\to b}^{(T)}(y,\mu^2)
D_{b\to H}\left(\frac{x}{y},y\mu^2\right),
\label{eq:f}
\end{equation}
which is to be solved imposing the boundary condition
$D_{a\to H}\left(x,m_H^2\right)=0$.
Evidently, the solutions of Eq.~(\ref{eq:f}) satisfy the condition
$D_{a\to H}(x,\mu^2)=0$ for $x<m_H^2/\mu^2$ as they should.
Notice that Eq.~(\ref{eq:f}) carries over to NLO as it stands.
One just needs to include the NLO corrections to $d_{a\to H}(x,\mu^2)$ and
$P_{a\to b}^{(T)}(y,\mu^2)$ and to employ the two-loop formula for the strong
coupling constant $\alpha_s(\mu^2)$.

\begin{figure}[ht]
\begin{center}
\centerline{\epsfig{figure=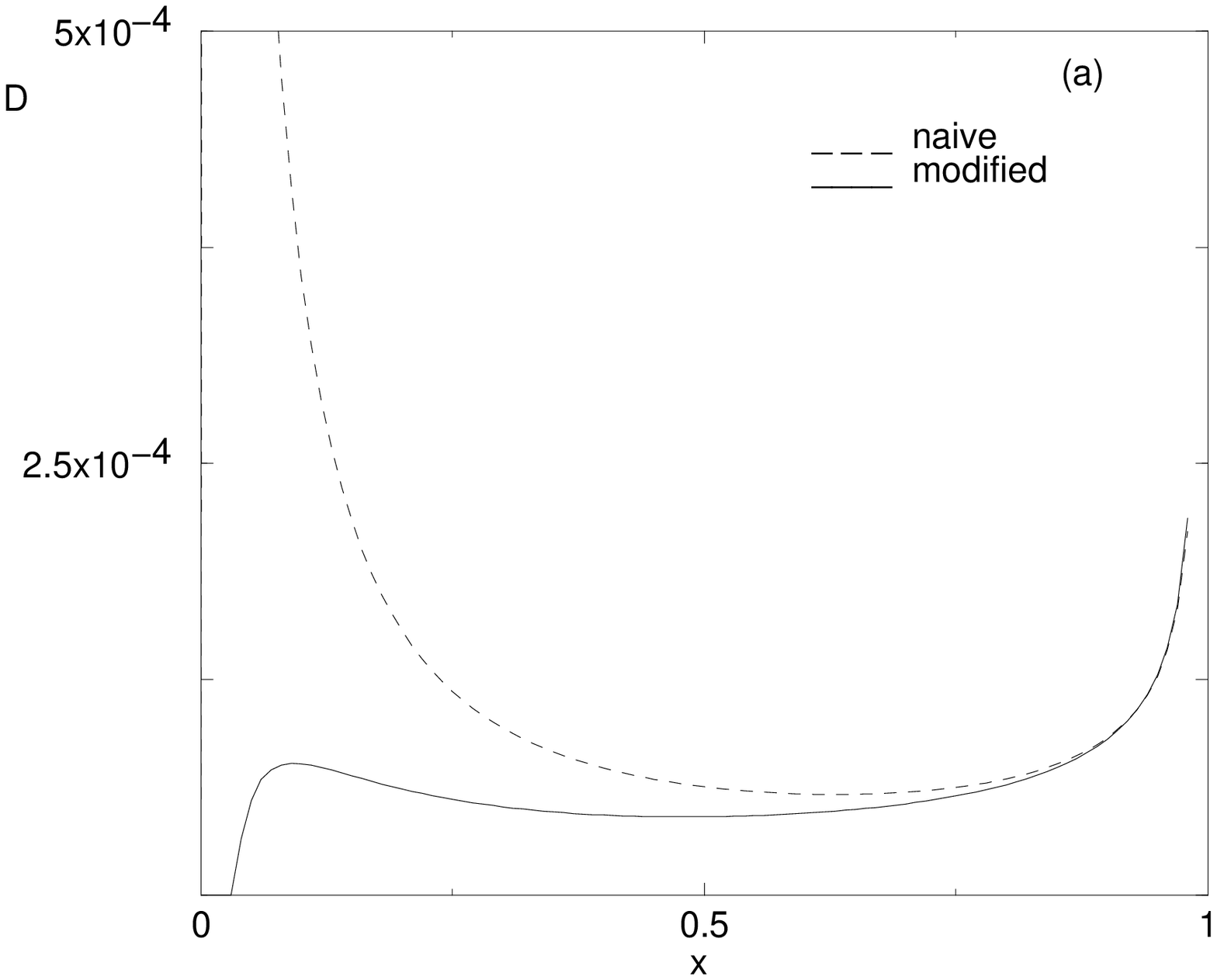,height=6.5cm}}
\centerline{\epsfig{figure=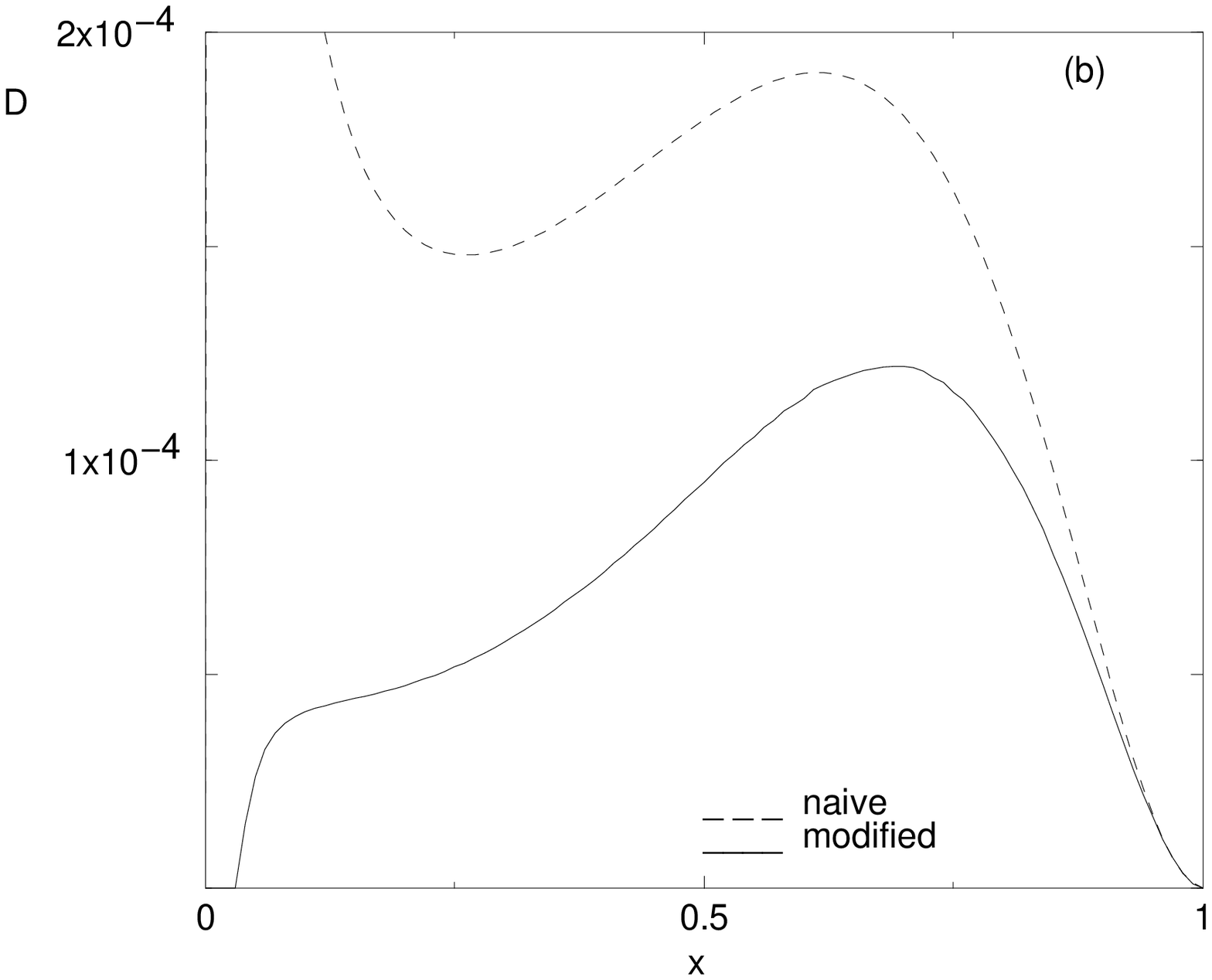,height=6.5cm}}
\centerline{\epsfig{figure=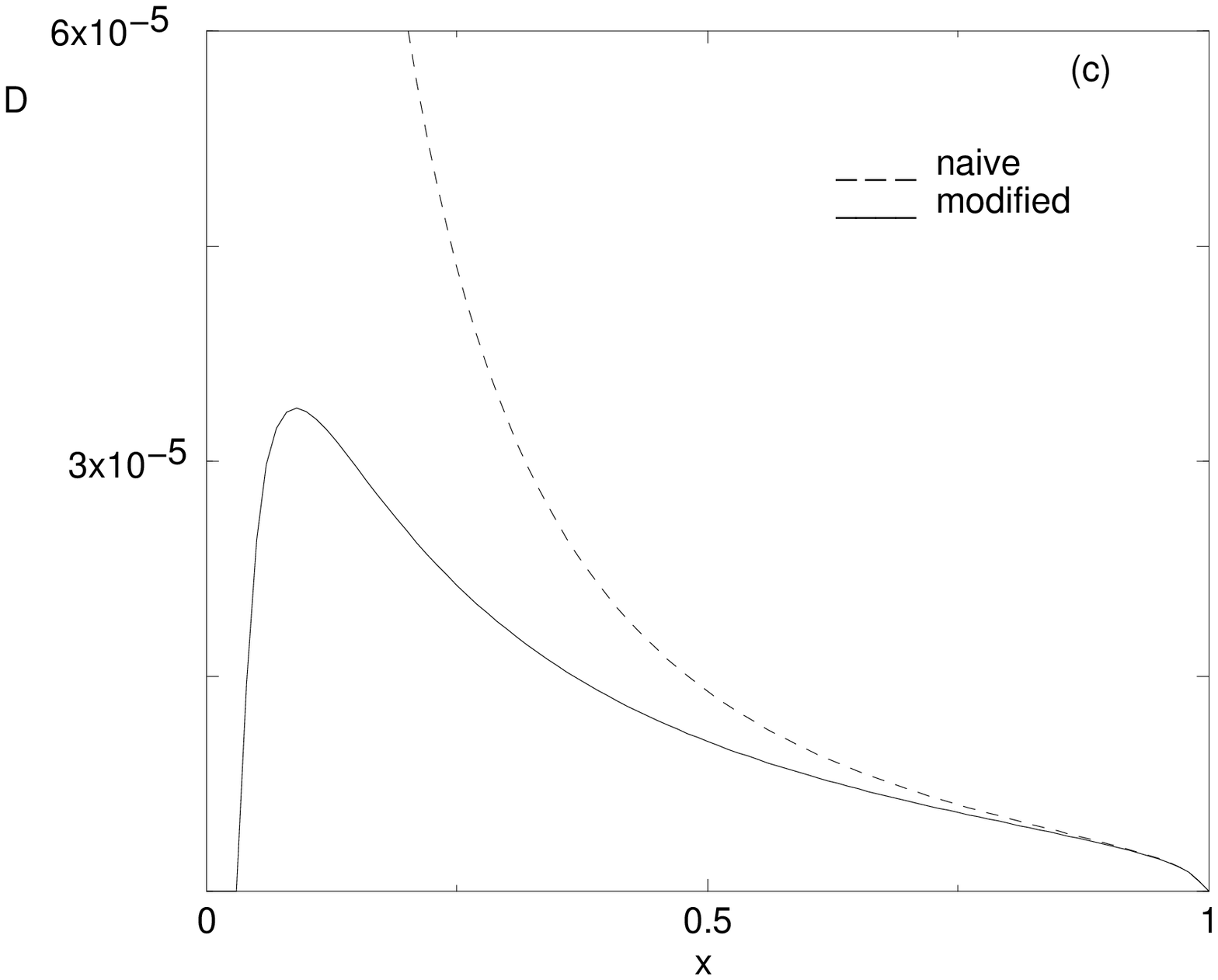,height=6.5cm}}
\caption{$D_{a\to J/\psi}(x,\mu^2)$ to LO at $\mu^2=300$~GeV$^2$ as a function
of $x$ for (a) $a=g$, (b) $a=c$, and (c) $a=u$.
The results obtained with the modified $\mu^2$ evolution (solid lines) are
compared with those obtained with the naive $\mu^2$ evolution (dashed lines).}
\label{fig:one}
\end{center}

\end{figure}

We now explore the phenomenological implications of the modified $\mu^2$
evolution for the case of prompt $J/\psi$-meson production via fragmentation,
$H=J/\psi$, in the framework of the factorization formalism of nonrelativistic
QCD (NRQCD) \cite{bod}.
At the starting scale $\mu_0$, the leading contributions arise from the
fragmentation processes $a\to c\bar c[n]$ with $a=g,c,\bar c$ and
$[n]=\left[\,\underline{1},{}^3\!S_1\right],
\left[\,\underline{8},{}^3\!S_1\right]$, where $\underline{1}$ and
$\underline{8}$ label colour-singlet and colour-octet states, respectively, and
the spectroscopic notation ${}^{2S+1}\!L_J$ indicates the spin $S$, the
orbital angular momentum $L$, and the total angular momentum $J$.
The corresponding fragmentation functions
$D_{a\to J/\psi}\left(x,\mu_0^2\right)$ may be found in Refs.~\cite{psi,kni} and
their source densities $d_{a\to J/\psi}(x,q^2)$ may be gleaned from
Ref.~\cite{psi}.
The fragmentation functions for $a=u,\bar u,d,\bar d,s,\bar s$ are generated via
the $\mu^2$ evolution and coincide.
Furthermore, we have $D_{c\to J/\psi}(x,\mu^2)=D_{\bar c\to J/\psi}(x,\mu^2)$,
so that we only need to distinguish the three cases $a=g,c,u$.
In Figs.~\ref{fig:one}a-c, we study the $x$ dependences of
$D_{a\to J/\psi}(x,\mu^2)$ at $\mu^2=300$~GeV$^2$ for $a=g,c,u$, respectively,
comparing the the naive (dashed lines) and modified (solid lines) $\mu^2$
evolutions to leading order (LO).
We adopt the LO color-singlet matrix element from Ref.~\cite{fit} and the LO
colour-octet ones from Table~\ref{tab:one}, selecting those which refer to the
naive $\mu^2$ evolution.
The $\mu^2$ evolutions are performed iteratively in $x$ space.
The technicalities are explained for the naive and modified $\mu^2$ evolutions
in Refs.~\cite{kni,zwi}, respectively.
As is well known \cite{bas}, the naive $\mu^2$ evolution equations~(\ref{eq:ap})
break down in the low-$x$ region due to the presence of large logarithms of
$1/x$ on their right-hand sides.
The most dramatic consequence of this deficiency is an unphysical divergence in
the gluon multiplicity at low $x$.
This directly triggers the singular low-$x$ behaviour of
$D_{g\to J/\psi}(x,\mu^2)$, which is exhibited by the dashed line in
Fig.~\ref{fig:one}a.
Via nondiagonal evolution effects, this feature also feeds into the quark
fragmentation functions, as may be seen from Figs.~\ref{fig:one}b and c.
By contrast, the results based on the modified $\mu^2$ evolution are devoid of
such low-$x$ divergences.
As per construction, they vanish for $x<m_{J/\psi}^2/\mu^2=0.032$.
On the other hand, they merge with the naive evaluation as $x$ approaches unity
because the impact of the phase-space constraint then fades out.  
Both features serve as a welcome check for our numerical analysis.
Further evidence for the physical soundness of the modified $\mu^2$ evolution
comes from a recent phenomenological analysis of $Z$-boson decay into prompt
$J/\psi$ mesons via fragmentation \cite{ern}, where the large logarithms of the
fixed-order calculation were resummed by using a Monte Carlo cascade model.
Except very close to the threshold region, the results thus generated were found
to nicely agree with those obtained from the modified $\mu^2$ evolution.

\begin{table}
\begin{center}
\caption{Values of the leading $J/\psi$ colour-octet matrix elements (in units
of $10^{-3}$~GeV$^3$) and of $r$ resulting from the LO and HO-improved fits to
the CDF data \protect\cite{abe} with naive and modified $\mu^2$ evolution.}
\label{tab:one}
\smallskip
\begin{tabular}{|c|c|c|c|c|}
\hline\hline
 & \multicolumn{2}{c|}{naive} & \multicolumn{2}{c|}{modified} \\
\cline{2-5}
 & LO & HO & LO & HO \\
\hline
$\left\langle{\cal O}^{J/\psi}[\,\underline{8},{}^3\!S_1]\right\rangle$ &
$4.58\pm 0.73$ & $2.58\pm 0.42$ & $5.01\pm 0.79$ & $2.86\pm 0.46$ \\
$M_r^{J/\psi}$ &
$104\pm 10 $ & $19.1\pm 2.8 $ & $99.9\pm 9.8$ & $17.0\pm 2.7$ \\
$r$ & 3.44 & 3.53 & 3.43 & 3.52 \\
$\chi_{\rm DF}^2$ & 0.52 & 0.52 & 0.52 & 0.41 \\
\hline\hline
\end{tabular}
\end{center}
\end{table}

At present, one of the most important applications of the $J/\psi$-meson
fragmentation functions is to describe the inclusive hadroproduction of prompt
$J/\psi$ mesons with high transverse momenta in $p\bar p$ collisions at the
Fermilab Tevatron, with $\sqrt s=1.8$~TeV.
In fact, fits \cite{fit,cho} to the latest data on $p\bar p\to J/\psi+X$ taken
by the CDF Collaboration \cite{abe} allow one to place stringent constraints on
the leading colour-octet matrix elements,
$\left\langle{\cal O}^{J/\psi}[\,\underline{8},{}^3\!S_1]\right\rangle$,
$\left\langle{\cal O}^{J/\psi}[\,\underline{8},{}^1\!S_0]\right\rangle$, and
$\left\langle{\cal O}^{J/\psi}[\,\underline{8},{}^3\!P_J]\right\rangle$, with
$J=0,1,2$, in the framework of the NRQCD factorization formalism \cite{bod}.
Specifically, the data in the upper $p_T$ range are especially sensitive to
$\left\langle{\cal O}^{J/\psi}[\,\underline{8},{}^3\!S_1]\right\rangle$, while
the low-$p_T$ data essentially fix the linear combination
\begin{equation}
M_r^{J/\psi}=
\left\langle{\cal O}^{J/\psi}[\,\underline{8},{}^1\!S_0]\right\rangle
+\frac{r}{m_c^2}
\left\langle{\cal O}^{J/\psi}[\,\underline{8},{}^3\!P_0]\right\rangle,
\label{eq:mr}
\end{equation}
where $r$ is to be chosen in such a way that the superposition of these two 
channels is insensitive to precisely how they are weighted relative to each 
other.
Due to heavy-quark spin symmetry, the multiplicity relation
$\left\langle{\cal O}^{J/\psi}[\,\underline{8},{}^3\!P_J]\right\rangle
=(2J+1)\left\langle{\cal O}^{J/\psi}[\,\underline{8},{}^3\!P_0]\right\rangle$ is
approximately satisfied.
Adhering to the analysis of Ref.~\cite{fit}, we now investigate by how much the
fit values for 
$\left\langle{\cal O}^{J/\psi}[\,\underline{8},{}^3\!S_1]\right\rangle$ and
$M_r$ are shifted if we pass from the naive to the modified $\mu^2$ evolution
The philosophy advocated in Ref.~\cite{fit} is adopt the fusion picture, where
the $c\bar c$ bound state is formed within the primary hard-scattering process,
in the low-$p_T$ regime and the fragmentation picture, where the $c\bar c$ bound
state is created from a single parton which is close to its mass shell, in the
high-$p_T$ regime.
We first redo the LO and higher-order-improved (HO) analyses of Ref.~\cite{fit},
which are based on slightly obsolete parton density functions, using the latest
LO and NLO MRST sets \cite{mrs}.
We then repeat these calculations with the modified $\mu^2$ evolution.
The resulting LO and HO values for
$\left\langle{\cal O}^{J/\psi}[\,\underline{8},{}^3\!S_1]\right\rangle$ and
$M_r^{J/\psi}$ are summarized in Table~\ref{tab:one} together with the
corresponding values of $\chi^2$ per degree of freedom, $\chi_{\rm DF}^2$.
The LO and HO values for
$\left\langle{\cal O}^{J/\psi}[\,\underline{1},{}^3\!S_1]\right\rangle$ remain
the same as in Ref.~\cite{fit}.
We observe that the effect of the modified $\mu^2$ evolution is to
increase $\left\langle{\cal O}^{J/\psi}[\,\underline{8},{}^3\!S_1]\right\rangle$
and to decrease $M_r^{J/\psi}$.
This may be understood from Figs.~\ref{fig:one}a--c by observing that the use of
the modified $\mu^2$ evolution leads to a reduction of the fragmentation
functions at low and intermediate values of $x$.
This  must be compensated by an increase of
$\left\langle{\cal O}^{J/\psi}[\,\underline{8},{}^3\!S_1]\right\rangle$ in order
to match the high-$p_T$ data, which are described in the fragmentation picture.
On the other hand, the use of this increased value of 
$\left\langle{\cal O}^{J/\psi}[\,\underline{8},{}^3\!S_1]\right\rangle$ in the
fusion picture must in turn be compensated by a reduction of $M_r^{J/\psi}$ in
order to fit the low-$p_T$ data.
Within each order of perturbation theory, the shifts in
$\left\langle{\cal O}^{J/\psi}[\,\underline{8},{}^3\!S_1]\right\rangle$
and $M_r^{J/\psi}$ due to the change of the evolution mode are relatively
modest, comparable to the errors on these quantities.
This may be understood by considering the average $x$ values which are probed by
$p\bar p\to J/\psi+X$ via fragmentation.
At the data point of highest $p_T$, with $p_T\approx18$~GeV, we have
$x=0.73\pm0.14$ in the colour-singlet channel, $x=0.90\pm0.12$ in the
colour-octet channel, and $x=0.90\pm0.12$ for their combination.
As is evident from Figs.~\ref{fig:one}a--c, the massive-evolution effect is not
yet pronounced at such high $x$ values.

In conclusion, the implementation of the phase-space constraint for heavy
hadrons in the $\mu^2$ evolution leads to a significant reduction of their
fragmentation functions at low and intermediate values of the
longitudinal-momentum fraction $x$.
However, current determinations of the leading colour-octet matrix elements for
the $J/\psi$ meson from Tevatron data of $p\bar p\to J/\psi+X$ are only modestly
affected by this because the $x$ values which are typically probed in this
reaction are rather high.

\vspace{1cm}
\noindent
{\bf Acknowledgements}
\smallskip

\noindent
One of us (B.A.K.) thanks the KEK Theory Division for the hospitality extended
to him during a visit when this paper was prepared.
The II. Institut f\"ur Theoretische Physik is supported by the
Bundesministerium f\"ur Bildung und Forschung under Contract No.\ 05~HT9GUA~3,
and by the European Commission through the Research Training Network
{\it Quantum Chromodynamics and the Deep Structure of Elementary Particles}
under Contract No.\ ERBFMRXCT980194.

\newpage

\end{document}